\documentclass[letterpaper,10pt]{article} 

\usepackage{opticameet3} 
\usepackage{multicol}

\newcommand\authormark[1]{\textsuperscript{#1}}

\usepackage{amsmath,amssymb}
\usepackage[colorlinks=true,bookmarks=false,citecolor=blue,urlcolor=blue]{hyperref} 

\begin{document}

\title{Simultaneous IM/DD Data Transmission\\ and High-Rate Secret Key Distribution\\
over a Single C-band Channel}


\author{M. Jachura,\authormark{1} J. Szlachetka,\authormark{2} M. Kucharczyk,\authormark{1} M. Jarzyna,\authormark{1} P. Kolenderski,\authormark{2} J.~P.~Turkiewicz,\authormark{3} and K. Banaszek\authormark{1*}}

\address{\authormark{1}Centre for Quantum Optical Technologies, University of Warsaw, ul.\ Banacha 2c, 02-097 Warsaw, Poland\\

\authormark{2}Institute of Physics, Nicolaus Copernicus University, ul.\ Grudziadzka 5, 87–100 Toru\'{n}, Poland\\

\authormark{3}Institute of Telecommunications, Warsaw University of Technology, ul.\ Nowowiejska 15/19, 00-665 Warsaw, Poland}

\email{\authormark{*}k.banaszek@uw.edu.pl} 

\begin{abstract}
We demonstrate hierarchical multiscale PAM-4 transmission combining 500 Mbps data transfer with optical-layer cryptographic key distribution at rates 24.14~Mbps and 8.38~Mbps secure against passive eavesdropper advantage 0~dB and 6~dB respectively. 
\end{abstract}

\section{Introduction}
Although the overwhelming majority of present-day optical communication systems relies on higher-layer protocols to ensure confidentiality and integrity of data transmission as well as user authentication, there is a growing recognition for the potential benefits of reinforcing information security with physical layer solutions
\cite{FokIEEETIFC2011,Skorin-KapovIEEECommMag2016}. Practical physical layer security (PLS) techniques need to be effective with respect to the required implementation overheads and ease of integration into existing communication systems. From this perspective, the recently proposed \cite{Ikuta_2016} and demonstrated in proof-of-principle setting \cite{Yamamori_2020} intensity modulation/direct detection (IM/DD) optical key distribution (OKD) offers an attractive option to realize one of basic PLS primitives, namely generating between two communicating parties a secure key (shared randomness) that will be unknown to an adversary with physical access to the communication channel. The security promise of IM/DD OKD covers passive eavesdropping, i.e.\ capturing fractional or entire part of the signal transmitted by the sender that is not received by the legitimate recipient, even if compared to the latter the eavesdropper has advantage in access to the signal \cite{Banaszek:21}. 

\begin{figure}[b]
  \centering
  \includegraphics[width=12.5cm]{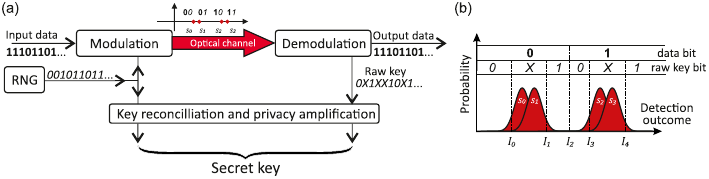}
\caption{(a) System architecture for simultaneous data transmission and optical key distribution using a hierarchical two-scale PAM-4 format. (b) Discrimination thresholds for detection outcomes applied to decode the transmitted data and to generate the raw key.\label{Fig:System}}
\end{figure}

In this contribution we demonstrate for the first time complete integration of the OKD protocol with regular IM/DD data transmission over a single  channel in the C-band transmission window. As shown in
Fig.~\ref{Fig:System}(a), this is accomplished by employing a hierarchical two-scale PAM-4 format where the coarse modulation scale is used to encode data, while the fine scale enables key generation. The detected signal is decoded using multiple-threshold discrimination depicted in
Fig.~\ref{Fig:System}(b) which provides both the data bit value and also, in postselected cases, raw key bits that are subsequently distilled to produce a secure cryptographic key. The implementation of the OKD protocol as an add-on to conventional IM/DD data transmission prospectively makes it a cost-effective PLS solution with minimal overheads in terms of the required additional bandwidth or signal power in scenarios where passive eavesdropping is the primary concern, such as free-space optical communications \cite{Lopez-MartinezIEEEPJ2015}, in particular satellite-to-ground optical links \cite{TrinhIEEETC2020,JachuraICSO2022}.

\begin{figure}[t]
  \centering
  \includegraphics[width=9.5cm]{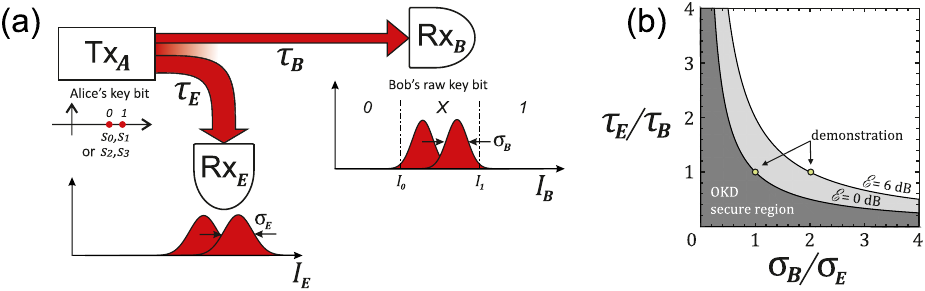}
\caption{(a) To generate a secure key, fine modulation depth at Alice's transmitter Tx$_A$ produces overlapping statistics of detection outcomes at Bob's and Eve's receivers Rx$_B$ and Rx$_E$. 
(b) The eavesdropper advantage ${\cal{E}} = (\tau_{E} /\sigma_{E})^2 / (\tau_{B} /\sigma_{B})^{2}$ is  used as a design parameter in the implementation of the OKD protocol that ensures security over a range of eavesdropping scenarios visualized as the shaded region. 
\label{Fig:KeySecurity}}
\end{figure}

\section{OKD protocol}

The key distribution protocol relies on fine modulation of the transmitted symbols by the sender Alice, who draws randomly their values from the pair either $s_0, s_1$ or $s_2, s_3$.
The choice between these two pairs is determined by the data bit to be transmitted (coarse modulation). 
As shown in Fig.~\ref{Fig:KeySecurity}(a), the fine modulation depth is chosen such that direct detection of the signal fractions $\tau_B$ received by the legitimate recipient Bob and $\tau_E$ captured by the eavesdropper Eve generates overlapping statistics of their detection outcomes, modelled as Gaussian distributions with respective widths $\sigma_B$ and $\sigma_E$. As the first step of
postprocessing, Bob sets discrimination thresholds to select outermost events outside the region marked as {\textsf X}, for which he can identify with a low probability of error $\varepsilon$ the key bit values used by Alice. The selected events, which form the raw key, are communicated by Bob to Alice over an authenticated public channel  without revealing obtained values. Provided that for a given symbol value Bob's and Eve's detection outcomes are statistically independent, Eve's knowledge about the selected string will be much less than that of either Alice's or Bob's, as her corresponding detection outcomes will likely fall into the intermediate region where inference of the key bit value is more difficult. 
Eavesdropper advantage in signal detection is compactly quantified by the ratio ${\cal E} = (\tau_{E} / \sigma_{E} )^2/ (\tau_{B} / \sigma_{B})^{2}$ that can be used as a design parameter to optimize Alice's fine modulation depth and Bob's discrimination thresholds, as well as to estimate Eve's knowledge about the raw key \cite{Banaszek:21}. 
The raw key needs to be distilled by Alice and Bob using communication over an authenticated public channel to correct for residual errors  and to remove Eve's knowledge (privacy amplification)
at the cost of shortening the final secure key. 
As illustrated in Fig.~\ref{Fig:KeySecurity}(b), a given value of the eavesdropper advantage ${\cal E}$ covers a range of eavesdropping scenarios with respect to the parameters of Bob's receiver subsystem. 
Importantly, Eve's detector performance is fundamentally limited by the unavoidable shot noise contribution, and the OKD protocol remains secure also when ${\cal E} > 1$ with the attainable key rate scaling approximately as ${\cal E}^{-1}$.

\section{Experimental setup and results}

\begin{figure}[htbp]
  \centering
  \includegraphics[width=13cm]{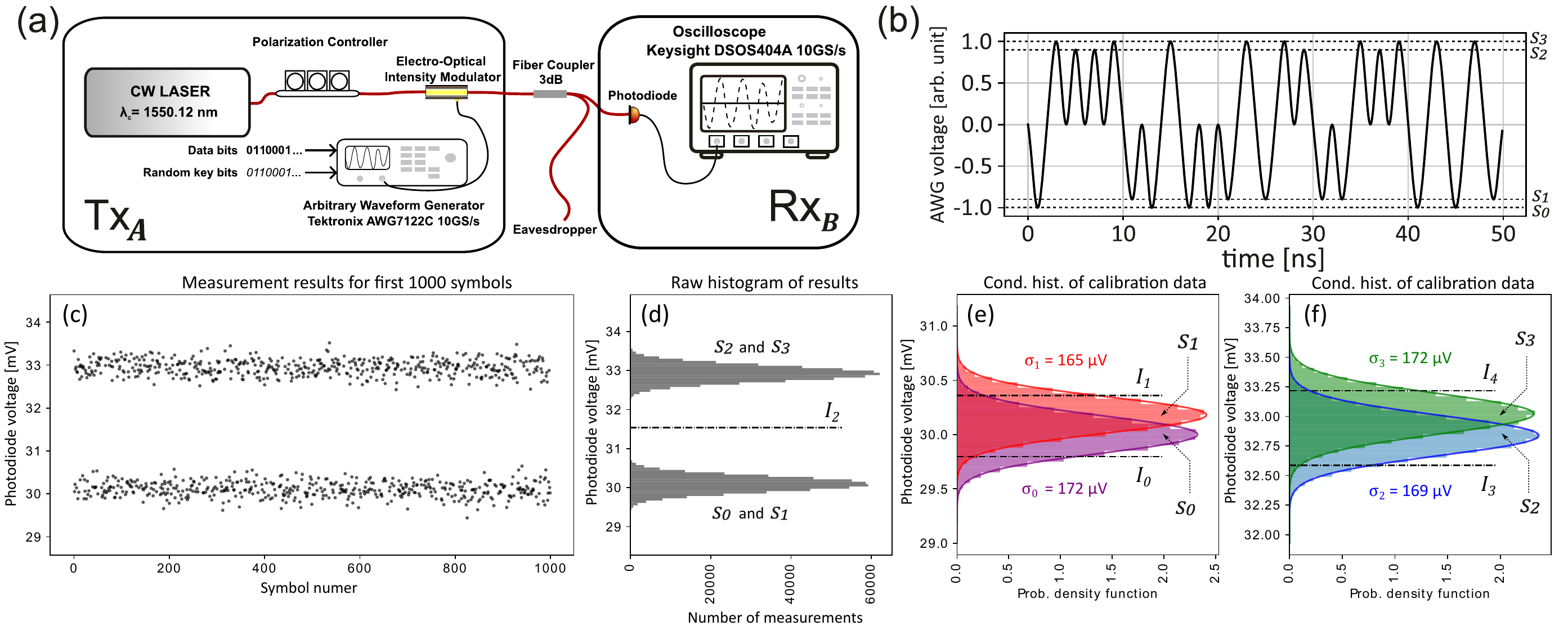}
\caption{(a) Experimental setup. (b) AWG-generated waveform with indicated symbol levels. (c) A sample of detection outcomes and (d) their histogram for the entire $10^{9}$ sequence. (e,f) Gaussian-fitted conditional histograms of detection outcomes for $10^8$ calibration symbols depending on the key bit  and the data bit values chosen by Alice. The oscilloscope noise floor was $153~\mathrm{\mu V}$.}
\label{Fig:Exp}
\end{figure}

The experimental setup shown in Fig.~\ref{Fig:Exp}(a) employed a $1550.12~\mathrm{nm}$ C-band CW-laser (ID Photonics CoBrite DX) to feed an electro-optic modulator driven by a $10~\mathrm{GS/s}$ arbitrary waveform generator (Tektronix AWG7122C). A pseudorandom binary sequence was utilized to generate the two-scale PAM-4 signal depicted in Fig.~\ref{Fig:Exp}(b) using a root-raised-cosine profile for the roll-off factor $0.6$ with 1~GHz slot rate and 1~ns guard time. The resulting effective symbol rate was 500~Mbaud. After 3~dB losses introduced by a 50/50 directional coupler to simulate eavesdropping the optical signal at approx.\ $1~\mathrm{mW}$ average power was measured with a photodiode (Optilab PD-23) directly connected to an oscilloscope (Keysight DSOS404A) with a sampling rate 10~GS/s. The arbitrary waveform generator and the oscilloscope were synchronized with a 10~MHz reference clock. 

In a single experimental run, a sequence of approx.\ $10^{9}$ symbols was transmitted with equiprobable occurrences of all four values $s_{0}, s_{1}, s_{2}$, and $s_{3}$. 
Fig.~\ref{Fig:Exp}(c) shows an exemplary segment of $10^3$  consecutive symbol readouts from the photodiode. The statistics of the entire $10^{9}$ sequence, depicted in Fig.~\ref{Fig:Exp}(d), exhibits two clearly separated peaks which enable decoding the data stream with the bit error rate (BER) at the level $9 \cdot 10^{-17}$ resulting from the Q-factor value of $8.24$. In order to generate a secure key, Bob used information about the key bit values utilized by Alice to modulate the first $10^8$ symbols in a sequence to resolve statistics generated by pairs $s_0, s_1$, and $s_2, s_3$ as shown in Fig.~\ref{Fig:Exp}(e,f) and to determine detection noise for individual symbol values. After characterizing noise, Bob processed the remaining approx.\ $9 \cdot 10^{8}$ events in the received sequence to obtain the raw key. The discrimination thresholds indicated in Fig.~\ref{Fig:Exp}(e,f) were selected to ensure equiprobable occurrences of $0$ and $1$ in the raw key and to keep the probability of error below $\epsilon \le 9.3\%$ for which a custom adapted LDPC code with a rate $1/2$ could be used for error correction. The resulting raw key was 123~752 bits long. The final key lengths obtained after applying the SHA3 privacy amplification algorithm assuming a given eavesdropper advantage are listed in Tab.~\ref{Tab:Results} along with the corresponding key rates.

\begin{table}[h]
\caption{Distilled key and the effective key rate depending on the assumed eavesdropper advantage.\label{Tab:Results}}
\centering
\begin{tabular}{|c|c|c|}
\hline
Eavesdropper advantage ${\cal E} $& Distilled key length & Key rate \\
\hline
\hline
$0~\mathrm{dB}$  &  43 454 bits& \textbf{24.14 Mbps} \\
\hline
$3~\mathrm{dB}$  & 31 120 bits & \textbf{17.29 Mbps} \\
\hline
$6~\mathrm{dB}$  & 15 080 bits & \textbf{\phantom{2}8.38 Mbps} \\
\hline
\end{tabular}
\end{table}

\section{Conclusions}

The presented results demonstrate that IM/DD data transmission and OKD can be efficiently combined into a single protocol using a hierarchical multiscale PAM format. While decoding the data stream follows the conventional routine, careful noise characterization and signal processing of the finely modulated signal enable simultaneous generation of a cryptographic key secure against passive eavesdropping. 
The demonstrated PLS solution shows many key advantages, including transmitter and receiver simplicity, high security level, and easy scalability to higher data rates. In space optical communication systems, it should facilitate conformity to current size, weight, and power consumption constraints of onboard communication terminals while adding robust protection against unauthorized signal access \cite{TrinhIEEETC2020,JachuraICSO2022}.




\bibliographystyle{opticajnl}  

\bibliography{sample}  

\end{document}